# Potentials of Unbalanced Complex Kinetics Observed in Market Time Series

Misako Takayasu[1], Takayuki Mizuno[1] and Hideki Takayasu[2]

[1]Department of Computational Intelligence & Systems Science, Interdisciplinary Graduate School of Science & Engineering, Tokyo Institute of Technology, 4259-G3-52 Nagatsuta-cho, Midori-ku, Yokohama 226-8502

[2]Sony Computer Science Laboratories, 3-14-13 Higashigotanda, Shinagawa-ku, Tokyo 141-0022

Abstract

As a model of market price, we introduce a new type of random walk in a moving potential which is approximated by a quadratic function with its center given by the moving average of its own trace. The properties of resulting random walks are similar to those of ordinary random walks for large time scales; however, their short time properties are approximated by abnormal diffusion with non-trivial exponents. A new data analysis method based on this model enables us to observe temporal changes of potential forces from high precision market data directly.

PACS    89.65.Gh   05.40.-a   05.45.Fb

In the late 18th century, Adam Smith applied the physics concept of force to market and introduced the concept of forces of demand and supply in economics, or "the invisible hands[1]." This classical mechanics model of market has been the standard base of modern economics; however, it is less known that the price dynamics has never been confirmed scientifically by real market data.

About 100 years ago Bachelier introduced a random walk model of market price[2] a little before the Einstein's famous paper on Brownian motion. It took nearly 70 years until his idea underwent a revaluation by the name of financial technology. In the formulation of financial technology the motion of market price is assumed to be a random walk without any market force. Although financial technology is now widely used in practical financial world, scientific validation is insufficient and deviations from real market data are pointed out[3,4].

Econophysics is a new frontier of science that is aimed to reconstruct economics by orthodox methodology of physics based on objective data analysis. Owing to the recent development of computer technology huge amount of detail market data are now stored and a lot of new empirical findings are established, such as power law distribution of price changes, long time volatility correlations and short time abnormal diffusions[5,6,7].

In this paper we introduce a new market price model that is a random walk in a moving and deforming potential function. The center of the potential is given by a moving average of the walker's traces. Based on this model we can observe the change of potential function directly from market time series. The market data we analyze here is the high-precision data in yen-dollar exchange market consisted of about 13 million bid prices in the period of 1995-2002.

The first step of our analysis is to introduce the optimal moving average that makes it possible to separate an independent noise from the raw data. For given time series of Yen-Dollar rate data, $\{p(t)\}$, the moving average is defined as

$$\overline{p(t)} = \sum_{k=1}^{n} w_k \, p(t-k) \tag{1}$$

where the weights $\{w_k\}$ are determined so that the residual term $f(t)$ defined by the following equation becomes an independent noise.

$$p(t) = \overline{p(t)} + f(t). \tag{2}$$

It is generally possible to satisfy $|\langle f(t) f(t+T) \rangle| < 1.0 \times 10^{-2}$ for $T$ = 1 to 1000 ticks, where the bracket denotes the average over $t$, and time is measured by a pseudo-time called the "tick time," that is the count of transaction numbers[8]. In general the estimated weights $\{w_k\}$ for the Yen-Dollar rates are roughly approximated by an exponential function $w_k \propto \exp(-0.3\,k)$, namely, the characteristic decay time is about 3 ticks which corresponds to around 30 seconds in real time on average.

For the motion of this moving average we assume the existence of a kind of linear central force with its center given by another moving average which we call as "the super moving average." This operation is defined by a simple moving average of past M ticks of the optimal moving average as follows:.

$$\overline{P_M(t)} \equiv \frac{1}{M}\sum_{k=1}^{M}\overline{P(t-k)}. \qquad (3)$$

Fig.1 shows an example of plots of raw market data, the optimal moving average and the super moving average. The value of $M$ is typically from 8 to 64.

The next step is to estimate the potential force by plotting the time difference of optimal moving average, $\overline{P(t+1)} - \overline{P(t)}$, versus the price difference between the optimal moving average and the super moving average at time $t$, $\overline{P(t)} - \overline{P_M(t)}$. If there is no force acting on the market as anticipated in financial technology, the plots should scatter around the horizontal axis. However, we can generally find non-trivial slope by this plot as typically shown in Fig.2 and the slope is proportional to the strength of the market force. In order to observe this force we use 2000 data points.

As this force can be regarded as a central force, we can calculate the potential function by integrating the force along the horizontal axis in the preceding figure. Fig.3(a) shows the examples of potential functions estimated for different values of $M$. The curvature or the strength of force is stronger for smaller $M$, and this $M$-dependence can be scaled by a factor $1/(M-1)$ and all curves collapse into one curve as shown in Fig.3(b). Thus, the dynamics of the optimal moving average derived from the data is given by the following equation including the effect of random motion that scatters the plots of Fig.2(a).

$$\overline{P(t+1)} - \overline{P(t)} = -\frac{1}{2} \cdot \frac{b(t)}{M-1} \cdot \frac{d}{d\overline{P}} (\overline{P(t)} - \overline{P_M(t)})^2 + F(t) \qquad (4)$$

Here, $F(t)$ is the residual term that is approximated by an independent random noise, and the coefficient $b(t)$ is estimated from the data independent of the bin size $M$ of the super moving average.

When $b(t)$ is positive the potential force attracts the optimal market price to the super moving averaged price, so the diffusion is slower than the case of no potential force. This tendency is confirmed by a numerical simulation of Eq.(4) with a constant $b$-value and independent random noise $F(t)$. Fig.4 (the bottom bold line) shows the time evolution of the mean deviation $\sigma(dt)$ for time difference $dt$ in log-log scale. For time scale shorter than 100 ticks, the diffusion can be characterized by a fractional power law consistent with the results of abnormal diffusion studies[9];

$$\sigma(dt) \propto (dt)^\alpha \qquad . \qquad (5)$$

However, for long time scale this abnormal diffusion property vanishes and the exponent of this random walk converges to that of a normal diffusion, $\alpha = 0.5$, namely, we have the following relation,

$$\sigma(dt) = \sigma_b \sqrt{dt} \qquad , \qquad (6)$$

where $\sigma_b$ is a time constant determined for each $b$-value.

In the case of negative $b(t)$ the potential force expels $\overline{P(t)}$ from $\overline{P_M(t)}$, and the resulting diffusion is faster than the normal diffusion as shown in Fig.4 (the top bold line). For short time scale, the abnormal diffusion characteristics can also be approximated by a non-trivial fractional relation Eq.(5), and its long time property is well approximated by a normal diffusion, Eq.(6).

The functional form of $\sigma_b$ is estimated both numerically and analytically[10] to be given by the following equation for the range, $-2 < b < 2$ (see Fig.5).

$$\frac{\sigma_b}{\sigma_0} = \frac{2}{2+b} \qquad (7)$$

As the diffusion constant $D$ is defined by $(\sigma(t))^2/t$ for large $t$, the $b$-dependence of $D$ is given by the square of Eq.(7). It should be noted that Eq.(7) diverges at $b = -2$. This result implies that in the case where $b \leq -2$ the potential force term in Eq.(4) dominates the random noise term $F(t)$ and we will have a monotonic dynamical price change in large time scale independent of the magnitude of the noise term.

The above theoretical results, from Eq.(5) to Eq.(7), are valid only for fixed value of $b(t)$, however, the observed value of $b(t)$ fluctuates in various time scales. Fig.6(a) shows an example of Yen-Dollar rate fluctuation and Fig.6(b) gives corresponding fluctuation of $b(t)$. It is confirmed that for the period of positive $b(t)$ the market fluctuation is actually stable, and for negative value of $b(t)$ the market fluctuates largely.

The observed autocorrelation function of $b(t)$ has the following long tail of power law for time scale up to about 3 months;

$$C_b(t) \propto t^{-0.2} \qquad . \qquad (8)$$

As the fluctuation of $b(t)$ is much slower than that of market rates, the abnormal diffusion properties for short time scale can be confirmed from the data as shown in Fig.7(a) and 7(b).

The model equation, Eq.(4), provides an excellent numerical simulation tool for markets. We compare direct observation of market data and a numerical simulation as for the distribution of exchange rate changes. In this simulation we apply the observed value of $\{b(t)\}$ and numerically created independent noise for $\{F(t)\}$. As shown in Fig.(8) the power law in the distribution is well re-produced by this simulation. Other basic properties can also be checked in a similar way[10]. From these results we can conclude that the time series of $\{b(t)\}$ contain the most important information about the market.

In this paper we succeeded in confirming the existence of market forces directly from the market time series by applying our new method named PUCK (Potentials of Unbalanced Complex Kinetics) analysis. There are two reasons why this type of attempts of observing market forces had been defeated so far. One reason is the amount and quality of data. We need about 500 to 2000 data points to estimate one value of $b(t)$ while its value changes even in a day from positive to negative. Namely, we need thousands of data points in a day, so the data interval must be shorter than a minute. This level of high frequency market data became available only in recent years.

The other reason is the lack of mathematical model. As we demonstrated, the underlying potential function is moving with its center given by the moving average of its own traces. Although there are huge amount of works on random walks in potential functions in the study of Ulenbeck-Ornstein process[11], there was no mathematical model in which the potential function moves according to its own traces. In actual markets the potential function is considered to be formed by the mass-psychology of dealers, so the position and strength of potential function should change in rather short time scale as dealers are generally believed to be quite sensitive to all kinds of information flows.

One of the most meaningful results in this paper is the existence of repulsive forces found in the market. This type of force has never been predicted so far, however, we believe that such unstable potential force is the key concept for understanding the whole behaviors of markets. We have a conjecture that the breakdown of diffusive motion for $b(t) \leq -2$ is related deeply to extraordinary market motions such as crashes, bubbles and hyper-inflations. This direction of study is now developed by our group intensively.

The method developed in this paper can be applied to any time sequential data.

We expect that potential functions can also be detected in apparently random data in various fields of science by applying the PUCK analysis.

This work is partly supported by Japan Society for the Promotion of Science, Grant-in-Aid for Scientific Research #16540346 (M.T.) and Research Fellowships for Young Scientists (T.M.). The authors appreciate H. Moriya of Oxford Financial Education Co Ltd. for providing the tick data.

**Figures**

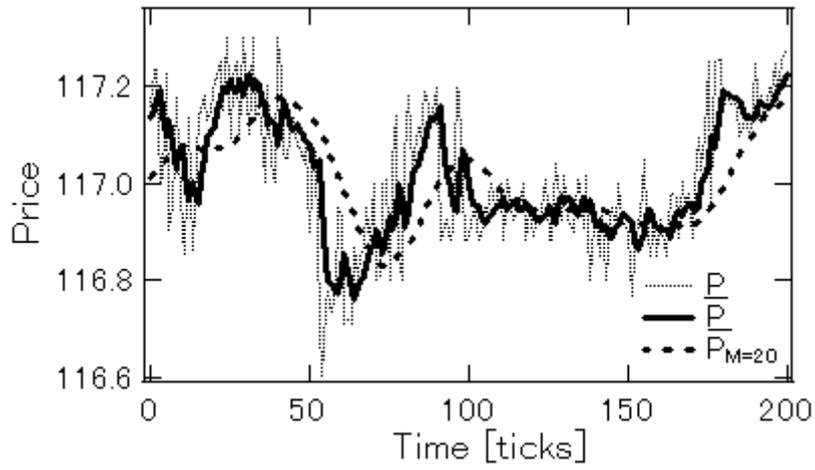

Fig.1 An example of raw data (dots), the optimal moving average (line) and the super-moving average with $M=20$ (dash).

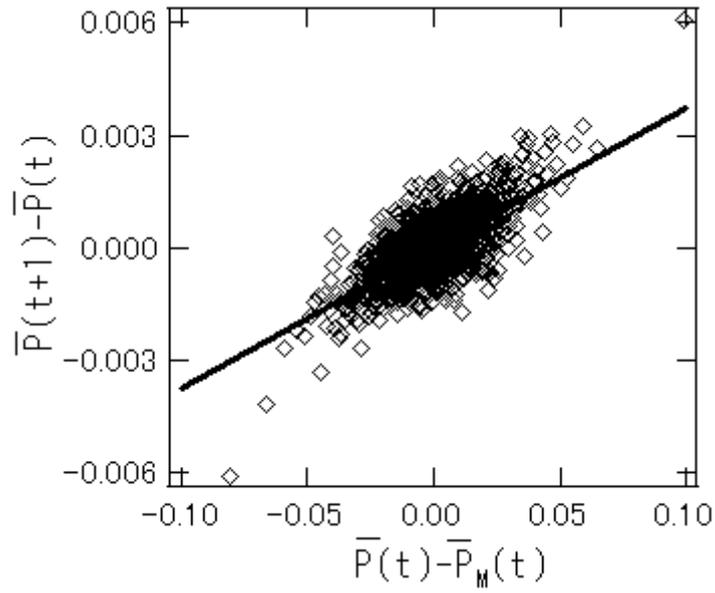

Fig.2 Time difference, $\overline{P(t+1)} - \overline{P(t)}$, versus the price difference, $\overline{P(t)} - \overline{P_M(t)}$, with $M$=16. The straight line denotes the least square fitting line for 100 data points.

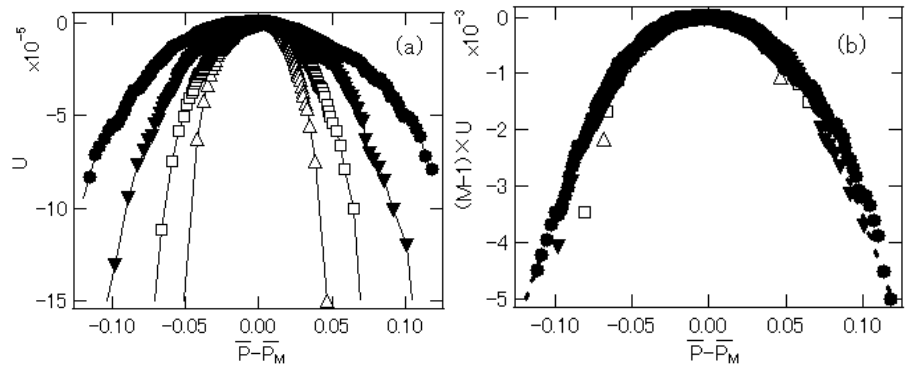

Fig.3(a) Potential functions $U$ for $M$=8, 16, 32 and 64 ticks.

Fig.3(b) Collapse of the potential functions by re-scaling.

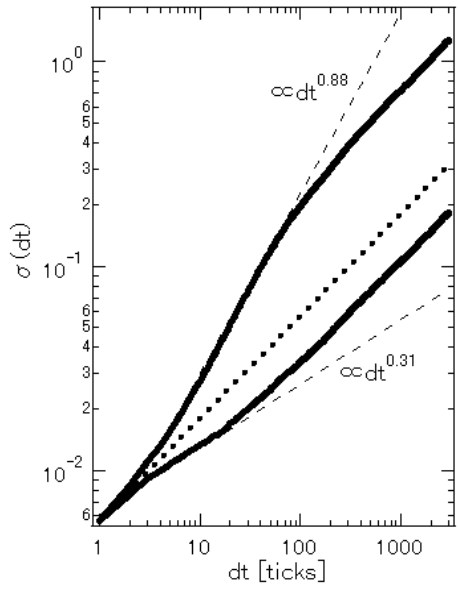

Fig.4 Theoretical estimation of the time evolution of the mean deviation in log-log plot. The dotted line shows the normal diffusion without any potential. The case of $b=1.5$ (bottom bold line) is approximated by an abnormal diffusion with $\alpha=0.31$ for short time scale. The case of $b(t)=-1.5$ (top bold line) gives the apparent slope with $\alpha=0.88$.

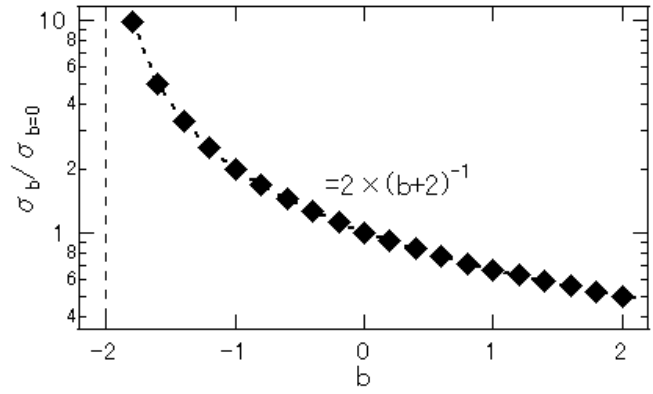

Fig.5 The functional form of $\sigma_b$. Theory (line) and numerical simulation (squares).

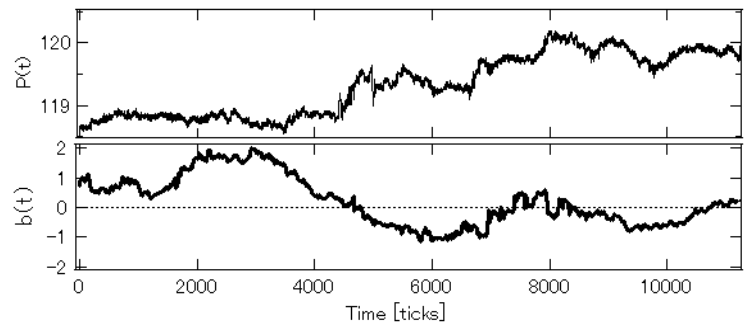

Fig.6 An example of one-day Yen-Dollar rate fluctuations on 7.3.2001 (top) and corresponding b-values (bottom).

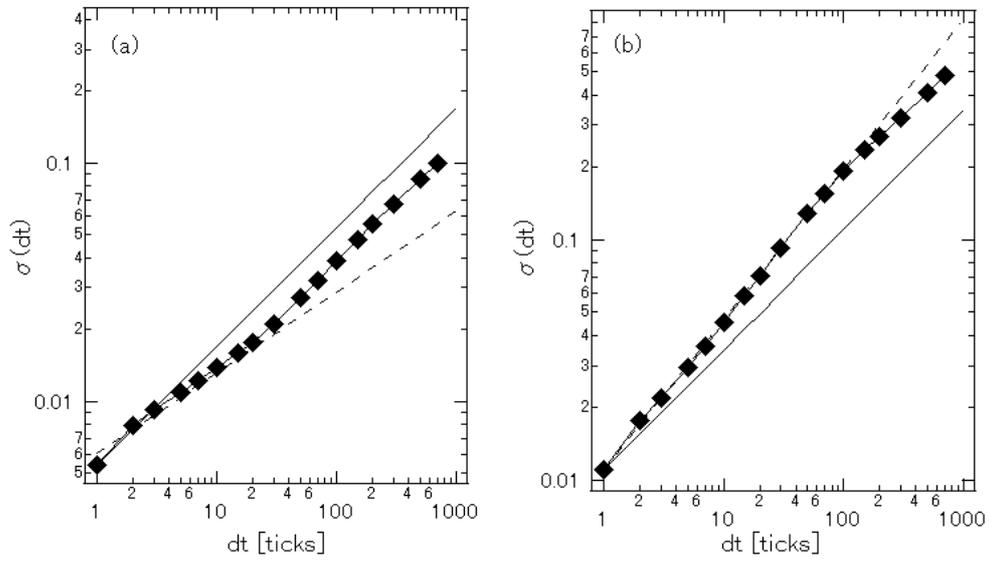

Fig.7(a) Observation of abnormal diffusion from real data, the value of $b(t)$ is about 1.6. The dotted line gives slope with $\alpha = 0.3$.

Fig.7(b) The case of $b(t)$ about $-1.2$. The dotted line gives slope with $\alpha = 0.7$.

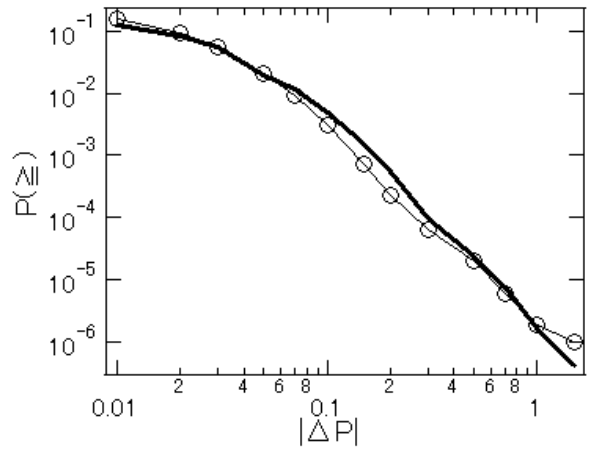

Fig.8 Distribution of rate changes $\Delta P = |P(t+1) - P(t)|$ in log-log plot. Real data (line) and the simulation using observed values of $\{b(t)\}$ and random numbers of $\{F(t)\}$.